\documentstyle[11pt]{article}

\begin{document}
\title{\bf
Plasmons in simple-metal slabs: a semiclassical approach.
}
\author{
{\bf
Sergio Leseduarte \thanks{lese@ecm.ub.es},
Jordi Sellar\`es \thanks{jordis@ecm.ub.es}
and Alex Travesset \thanks{alex@greta.ecm.ub.es}
\thanks{Authors listed in alphabetical order}
}\\
Department d'Estructura i Constituents de la Mat\`eria,\\
Universitat de Barcelona, Diagonal 647,\\
E-08028 Barcelona, Catalonia}
\maketitle
\begin{abstract}
Collective excitations in simple metal systems can be described
successfully in terms of a local one-body excitation operator Q,
due to the long range nature of the coulomb interaction.
For the plasmon modes of a simple-metal slab, momentum
expansions of Q are calculated using 
a variational procedure, equivalent
to a restricted RPA calculation. The 
dispersion relation and the density
fluctuation for each mode are found 
in the sudden approximation using
the proper Q operator and the RPA sum-rule 
formalism. The contribution of the
exchange and correlation energy is estimated using a 
local density functional.
The positive background is described 
within a jellium model while the
ground-state electronic density is approximated 
by a double step profile.
The density fluctuation of the plasmon modes 
above the plasma frequency form
standing waves across the slab.
The spectra below the plasma frequency is 
qualitatively different to that
of local optics calculations, due to the 
appearance of two multipole plasmon
modes that shift down the origin of
the $\omega_{+}$ plasmon. The dependence of the results
on the width of the slab, the density of the 
simple-metal and the surface
diffuseness is discussed. Throughout, we 
compare our results with previous
RPA and TDLDA calculations.
\end{abstract}

\vfill
\vbox{
 UB-ECM-PF-96/06\hfil\null\par
 March 1996\hfil\null\par
 \hfil\null\par
 \hfil\null\par}

\newpage

\section{Introduction.}
There are several ways in which the dispersion relation of the plasmon
modes in systems with translational invariance can be measured.
The first experiments on simple metal slabs
used a dielectric substrate to support the metal film and measured the
light transmittance \cite{1} or the photoelectric yield \cite{2}. More
recently, molecular-beam epitaxy has been applied to grow thin metallic
layers between two insulators \cite{3} or $Al_{x}Ga_{1-x}As$ heterostructures
simulating a positive jellium background \cite{4}. Also electron energy
loss spectroscopy (EELS) has been used to study adsorbed simple-metal layers
on an aluminium substrate \cite{5}.

The spectrum of a simple-metal slab has two distinct parts: the plasmon
modes that lie above $\omega _{p}$, corresponding to standing
waves across the slab, and the surface plasmon modes, with lower frequencies.
Some theoretical calculations have been done using the
random-phase approximation (RPA) \cite{6,7,8,9,10} and the time-dependent
local density approximation (TDLDA) \cite{11,12,13}. The $\omega_{-}$
and the $\omega_{+}$ plasmons are studied thoroughly in all this
references, but only two of them
discuss the existence of an additional multipole plasmon mode.
Liebsch \cite{12}, in a calculation of the plasmon modes of a simple-metal
overlayer on an aluminium substrate, finds a multipole plasmon
that hybridizes with the $\omega_{+}$ plasmon. Schaich and Dobson \cite{13}
also obtain a peak in the response function at frequencies
between $\omega_{p}$ and
$\omega_{p}/\sqrt{2}$ but they do not conclude whether the cause is a
collective mode or is due to intersubband excitations.

This multipole plasmon mode has
been detected in simple-metal clean surfaces using EELS \cite{14}. Its
dispersion relation has a positive slope when the momentum $k$ parallel
to the surface is zero while the frequency at this point is $0.8 \omega_{p}$.
This value
agrees with theoretical calculations \cite{14,15}.
Multipole plasmon modes are associated with electronic-density fluctuations
that are peaked at the surface region and have decreasing oscillating
amplitude towards the interior of the metal. It can be shown \cite{16} that
the integral of the electronic density perpendicular to the surface is zero.
Unlike ordinary surface plasmons, they are optically active because they
carry momentum on the normal direction.
All this properties were found for simple-metal surfaces but one expects that at
least some of them should be valid for slabs.

The aim of this work is to study the existence of multipole plasmon modes
on simple-metal slabs and its coupling to the other surface modes. We will use
the RPA sum-rule formalism (SR) that previously has been able to produce useful
results for different geometries \cite{17,18,19,20}. In particular, a
good agreement has been obtained between experimental data and the
dispersion relation of surface plasmons \cite{21} and multipole
plasmons \cite{15} on a plane simple-metal surface. It has
some advantages over other methods. The different contributions to the energy
of each mode can be analyzed on its own and semiclassical approximations for
the kinetic energy are easily implemented.
It can incorporate electronic exchange and correlation interactions using
a local density approximation. Within the SR formalism it is not difficult to
employ realistic electronic profiles. In fact the results have RPA precision
as long as a self-consistent ground-state electronic profile is used \cite{22},
although only an average of the response function can be obtained.

Usually, when working within the SR method, it is assumed that
for every collective
mode exists a local one-body observable that excites that mode without mixing
it with others. In order to obtain useful results an appropriate excitation
operator $Q$ must be found and the system must have well-defined collective
modes. Otherwise it would not be possible to get any valuable
information from the averaged response function. There is not a unique way
to find such an operator. Sometimes the properties of a given operator $Q$
determine the kind of mode that it excites.
For example, an operator that fulfills Laplace's equation can only excite
surface modes \cite{21}.
A restricted variation of $Q$ on a parameter
\cite{15} can be useful if a sufficiently general form of $Q$ is employed.
In RPA calculations an excitation operator $Q$ can be found for each mode
but although it will be an observable, in general it is not a local operator.
Nevertheless, the fact that the interaction responsible for the existence
of well-defined collective excitations, the coulomb interaction, has a
long range, allows
to approximate $Q$ as a local operator. This approximation is called
local-RPA \cite{23} and has been applied successfully to nuclei \cite{24} and
metal clusters \cite{25}. We use local-RPA to find a proper $Q$ operator as a
momentum expansion.

Instead of a self-consistent electronic profile, a double step profile
is used to describe the ground-state. This leads to a simplification
of the calculations which can be performed analytically without a great loss
of precision because it has been shown
\cite{21} that there is little quantitative difference 
between the results from
any of the two models.

\section{Application of local RPA to a metal slab}

In this section we first give a short review of the method as it was
stated in \cite{23} (see this reference for more details). Then we
expatiate on the particularities that we have had to devise to adapt it
to our problem, the most important being the choice of the form of the
functions that give rise to the sought-after excitation operators.

Both local and full RPA demand as a starting point a description of the
vacuum of the system at issue. Within the quasi-bosonic
approximation this description
is furnished by working out the Hartree-Fock approximation to the vacuum
state $| \phi_0 \rangle$.
This approximation is acceptable as long as
one assumes that correlations are
small. Once the Hartree-Fock problem is solved, the essence of the
method is to build the excitation operators as linear combinations of
one-particle--one-hole operators, and determine the coefficients of the
expansion by solving the equation of motion
\begin{equation}
\langle \phi_0 | \left[ C_m , [ H , C_n^{\dagger} ] \right] |
\phi_0 \rangle = \delta_{n m} \hbar \omega_m ,
\end{equation}
where $C_n^{\dagger }$ is an excitation operator. By this name 
we mean that when it acts on the ground state, an 
elementary excitation of
the system with a well defined energy is created.
An alternative description of RPA can be made based upon the operator
\begin{equation}
Q_n=\frac{C_n^{\dagger}+C_n}{\sqrt2}
\label{qopera}
\end{equation}
\begin{equation}
\left[H,\left[H,Q_n\right]_{ph}\right]_{ph}=
{\left(\hbar \omega_n\right)^2Q_n},
\label{paho}
\end{equation}
The $ph$ symbol means that the operator is
projected onto the linear space spanned by
one-particle--one-hole states.
Local RPA assumes that the operator $Q_n$ can be expanded
in terms of a function $f_n$ as follows,
\begin{equation}
Q_n = \sum_{i=1}^N f_n(\hat{r}_i)\,,
\label{localdef}
\end{equation}
where $N$ is the number of electrons and $\hat{r}_i = (x_i\,,
y_i\,,z_i)$ is the position operator of the i-th particle.
In order to determine $f_n$ we shall consider that it is a linear
combination of the elements of a basis of functions that we shall choose
taking advantage of the particular problem we are facing at.
We will focus on the problem of a gas of electrons in a simple metal
slab of very large area $A$ and thickness $d$. If the gas is made of $N$
electrons, then the operator of total momentum $q_{tot}$ of the gas will
be;
\begin{equation}
{\hat q}_{tot}=\sum_{i=1}^N {\hat q}_i
\end{equation}
We will assume the slab to be parallel to the $XY$ plane and
consequently write any 
vector as $q=(k,p)$, $k$ being the 2
dimensional
vector that lives in the $x-y$ plane and $p$ is the projection along the
$z$ axis. Translation invariance in the $x$ and $y$ axis leads to the
conservation law;
\begin{equation}
\left[H,{\hat k}_{tot}\right]=0
\end{equation}
So, the eigenvalues $k$ of operator ${\hat k}$ are good quantum numbers
for our system.
As $k$ is a true quantum number of the problem, we may advance
that each $Q_n$ operator has a well defined $k$, in other words,
it has the form
\begin{equation}
Q_{n}=\sum_{j=1}^{N} e^{ik\hat{R_{j}}} \phi_{n}(\hat{z_{j}}).
\label{cansat}
\end{equation}
This observation suggests that we 
relabel $Q_n$ as $Q_{k,n}$
in the sequel.

As for the $f_n$ functions, which
we haste to relabel in accordance
with the previous discussion, we may write
\begin{equation}
f_{k,n}(R,z)=\exp{(ikR)} \phi_{n}(z).
\label{cansat_cansat}
\end{equation}
The electronic profile vanishes for distances far enough
from the boundaries of the slab. We can think the gas of
electrons to live in a linear box of size L. We must take
L to be larger than d as there are electrons wandering
outside the slab, but provided L is large enough, the
physical results are L-independent. Nevertheless the
value of L must not be larger than necessary because this
would increase the number of terms needed in the expansion.

Now we have a physically motivated choice for the basis
which generates $\phi_{n}$; we take $\phi_{n}$ to be
expanded in terms of a Fourier basis
\begin{equation}
\phi_{n}(z)=\sum_{l} a_{n}^{l} e^{ip_{l}z},
\label{sempre_cansat}
\end{equation}
with $p_l = \frac{2 \pi l}{L}$. Equivalently, we may expand
the $Q_{k,n}$ operators
\begin{equation}
Q_{k,n}=\sum_{l} a_{n}^{l} Q_{k,p_{l}},
\label{tinc_idees}
\end{equation}
with
\begin{equation}
Q_{k,p_{l}}=\sum_{j=1}^{N} \exp{(ik\hat{R_{j}})} \exp{(ip_{l}\hat{z_{j}})}.
\end{equation}
By construction, the $Q$ operator will have a period L. This is done
just for mathematical convenience.
We have truncated the previous Fourier series ((\ref{sempre_cansat}) or
(\ref{tinc_idees})) making sure that the terms we do not
consider are really negligible. This can be done checking the convergence
of the results as the number of terms is increased.
It is clear that $L$ cannot be too small
as we would cut the tails of the electronic density, but it should
not be larger than necessary because this would increase the number
of terms needed in the preceding expansion.
$a_{n}^{p}$ are coefficients that can be obtained by minimizing
the energy
functional \cite{23}. Written in terms
of the Fourier coefficients eq. \ref{paho} reads 
\begin{equation}
\left[{\cal K_{\alpha \beta}}-(\hbar \omega_n)^2{\cal B_{\alpha \beta}
}\right] a_n^{\alpha} = 0,
\label{eigen}
\end{equation}
$\cal K_{\alpha \beta}$ and $\cal B_{\alpha \beta}$ being respectively;
\begin{equation}
{\cal B_{\alpha \beta}}=\langle \Phi_0 |\left[Q_{\alpha},\left[H,
Q_{\beta} \right]\right] | \Phi_0 \rangle,
\end{equation}
\begin{equation}
{\cal K_{\alpha \beta}}=\langle \Phi_0 | \left[\left[Q_{\alpha},H
\right]
,\left[ H,\left[ H,Q_{\beta} \right] \right] \right] | \Phi_0 \rangle.
\label{calK}
\end{equation}
$H$ is the Hamiltonian of the system and $| \Phi_0 \rangle $ the
(Hartree-Fock) ground state. Needless to say that the sum-rule
approach
\begin{equation}
m_{1}={1\over{2}}\langle \Phi_0 |\left[Q,\left[H,
Q \right]\right] | \Phi_0 \rangle,
\label{SRM1}
\end{equation}
\begin{equation}
m_{3}={1\over{2}}\langle \Phi_0 | \left[\left[Q,H
\right]
,\left[ H,\left[ H,Q \right] \right] \right] | \Phi_0 \rangle,
\label{SRM3}
\end{equation}
\begin{equation}
E_{3}=\sqrt{m_{3}\over{m_{1}}},
\label{SRE3}
\end{equation}
is a particular case of eq. \ref{eigen}
because if we could find a
series expansion such that every term would couple to a different
eigenmode, eq. \ref{eigen} would be separated into a set of
equations like eq. \ref{SRE3}, one for each term of the series.
It can be shown \cite{27} that $E_3$ is an upper bound to the
energy of the mode excited by $Q$.

Solving equation \ref{eigen} for a fixed value of $k$ does not only
deliver the optimal operator $Q$ but also the dispersion relations
of the collective excitations $\omega_n(k)$.
We calculate the dispersion relation in an equivalent, but more
transparent way. As the Hamiltonian
\begin{eqnarray}
H&=&\int{dr \left[ \frac{{\hbar}^2}{2m_{e}}\tau(r)+
\frac{n(r)}{2}\left(e^2\int{dr'\frac{n(r')}{\|r-r'
\|}}+2v_j(r)\right)\right]}
\nonumber
\\
&\equiv& H_k+ H_c.
\label{hamiltonian}
\end{eqnarray}
contains a kinetic and a coulomb term, we can evaluate separately
the contribution of each part of the Hamiltonian to the $m_3$
sum-rule. When the $Q$ operator used is that obtained solving
eq. \ref{eigen}, the eigenvalues of this equation are, by
construction, the $E_3$ energy obtained from eq. \ref{SRE3}
taking into account all the contributions to $m_3$.
If $E_3$ is calculated using only a contribution to $m_3$,
the quantity obtained gives an idea of the relative importance
of that term in the total energy. The inclusion of an
exchange and correlation term when solving eq. \ref{eigen}
is done in a perturbative way, as explained later on.

Another important magnitude in the sum-rule formalism is
the sudden approximation to the density fluctuation given
by
\begin{equation}
n_1=-\hbar \vec{\nabla}\left[ n \vec{\nabla} Q \right].
\label{SRN1}
\end{equation}
It is the first-order contribution to the density fluctuation of
the $|\eta\rangle$ state described in the appendix and usually
gives a reliable description of the eigenstates of the system.

At this point we just need to solve equation \ref{eigen} and we will do
so by expressing $\cal K_{\alpha \beta}$
and $\cal B_{\alpha \beta}$ in terms of the local density $n(r)$ and
the kinetic-energy density $\tau(r)$ (due to the symmetry discussed so
far the only dependence of this quantities is on the $z$ axis)
\begin{equation}
n(r)=\sum_{i}{|\phi(r)|}^2, \qquad \qquad
\tau(r)=\sum_{i}{|\nabla\phi(r)|}^2,
\end{equation}
To compute the kinetic term the most direct
approach consists in taking the kinetic energy of a free gas of
electrons;
\begin{equation}
\tau=an(r)^{5/3}.
\label{density}
\end{equation}
The coulomb part contains the coulomb interaction between electrons
(of charge $e$) and the interaction
with the jellium through a potential
$v(r)$.
An estimation of the exchange and correlation energy is made
using the following Slater- and Wigner-type expressions
\begin{equation}
\epsilon_{ex}=-\frac{3}{4}an(r)^{1/3}-b\frac{n(r)^{1/3}}{cn(r)^{1/3}+d}
\end{equation}
with $a=\sqrt{3/{\pi}}$,$b=.44$,$c=7.8$ and $d=a/2$.
Although this term is not included in the Hamiltonian when computing
$Q$, a contribution to the $m_3$ sum-rule is calculated just as if
it were another part of eq. \ref{hamiltonian} and added to the other
contributions. So the final values of the dispersion relation
are slightly different to the ones obtained in eq. \ref{eigen}.
In this way the exchange and correlation interactions are
included in the results. The main drawback is that the
self-consistency of the calculation is lost to a certain extent
but this is acceptable because exchange and correlation do not
play a crucial role in the existence of the collective
excitations in an electron gas.

The details necessary to solve equation
\ref{eigen} and improvements to the kinetic energy functional are
postponed to the appendix as they become rather technical and do not
contain any new physics.

\section{Results and discussion}
We have applied the method described in the preceding section to
simple metal slabs.
A double step electronic profile is used as a description of the
ground state
\begin{equation}
n(z)={n_b \over 2} \left[ \theta(z+{{d+\delta} \over 2})
+ \theta(z+{{d-\delta} \over 2}) -
\theta(z-{{d+\delta} \over 2}) -
\theta(z-{{d-\delta} \over 2}) \right].
\label{profile}
\end{equation}
The length of the step $\delta$ has been adjusted in all the cases
by a minimal squares procedure to an improved Thomas-Fermi electronic
profile \cite{26} calculated for each system.
In this section atomic units are used throughout.

Figure 1 (a) represents the dispersion relation
of the plasmon modes below $\omega_{p}$ for a typical system. The lower
mode is the $\omega_{-}$ mode.
It begins at $\omega=0$ for $k=0$
and tends to the surface plasmon frequency $\omega_{p}/\sqrt{2}$ as
$k$ tends to infinite. This behaviour is completely in agreement
with Local Optics, not like the next mode, that
we identify as the
$\omega_{+}$ mode of Local Optics. In our
calculation it has a frequency of approximately $\omega=0.8\omega_{p}$
for $k=0$ when according to Local Optics, and most RPA and TDLDA
calculations, it should tend to $\omega=\omega_{p}$ for a small $k$.
For large values of $k$ it becomes degenerate with the $\omega_{-}$
mode. There is another mode whose frequency tends to $0.8\omega_{p}$
for a low $k$. We call it the even multipolar mode and it has a
qualitative behaviour very similar to that of multipolar modes in a
surface. Its dispersion relation has a positive slope when $k=0$ and
tends to the same value as in the semi-infinite case. In
our calculation it becomes degenerate for large $k$
with a mode that we call the odd multipolar mode. This last mode has its
origin at the plasma frequency $\omega_{p}$. Although this scheme is
not the usual one in semiclassical or fully quantum-mechanical
calculations of collective excitations in metal slabs, a nearly
identical behaviour has been reported for plasmons in a simple-metal
overlayer \cite{12}. Figure 1 (b) displays the dispersion relation of
the first four bulk resonances. These dispersion relations are
qualitatively similar to those obtained by discretization of the $z$
component of the momentum
in a bulk plasmon dispersion relation but the slopes
are flatter than those obtained with other methods, even classical
\cite{2}. Also the position of the origins does not seem
to be quadratic as a reasoning of this type would indicate.

Further insight can be gained looking at the sudden approximation of
the density fluctuation for each mode.
Every one of the four discontinuities in the
ground-state electronic profile produces an avoidable discontinuity
in the density fluctuation. In those points the first derivative of
the density fluctuation is also discontinue. A more realistic
profile should be used to obtain quantitatively correct results
but as the double-step model includes the most important physical
aspects, such as diffuseness, qualitative information like symmetry
and number of nodes is reliable.
The density fluctuation of the $\omega_{-}$ and the $\omega_{+}$
modes is shown in Fig. 2 (a). The symmetry and the number of
nodes of this modes are the same as those of the even multipole
and the odd multipole respectively, in Fig. 2 (b). All these modes are
clearly surface modes and the main difference
between them is that in the
multipole modes the density fluctuation is not constant in the
middle of the slab. The decreasing oscillating amplitude that
multipole plasmons show in a semi-infinite medium is not found
in our calculation for thin slabs.
The density fluctuation
of the bulk resonances is represented in Fig. 2 (c)-2 (d).
The first resonance has two nodes and presents the typical profile of a
standing wave across the slab. In fact, we can label a resonance
with its number of nodes because any resonance has one
more node than the previous one.

The key part of the calculation is to choose the number $M$ of terms of
the $Q$ expansion. This is done, for a given momentum $k$ parallel to
the surface, examining the energy of the modes in front of $M$. The value
of $M$ necessary to attain the convergence depends on a great number of
physical factors. As the main assumption of Local RPA is the existence
of well defined collective excitations, the convergence will be
difficult or even impossible in all those situations where a severe
damping of the plasmon modes is expected. Because the relative importance
of the coulomb energy increases with the electronic density,
in high-density
metals like aluminium the convergence will need lower values of $M$
than low-density metals like cesium. Also when the
momentum $k$ approaches the Landau cutoff value, given approximately by
$q_{c}=\omega_{p}/v_{F}$, the convergence must be harder. The
surface diffuseness also plays its role because an increasing diffuseness
tends to lower the coulomb energy. 
All these general trends are present
in our calculations. The size of the system is also an important factor.
This is mainly because the expansion series of a bigger system needs
more terms to be accurate. This puts un upper bound on the size of
the slabs that can be treated within this method.

Taking advantage of the transparency of the method, the contributions
of the kinetic and the coulomb parts of the hamiltonian to the
dispersion relation are represented in Fig. 3.
The ratio between the coulomb and the kinetic energy
is different for each mode. The $\omega_{-}$ mode
has a negligible kinetic energy in front of the coulomb contribution while
they are comparable in the $\omega_{+}$ mode. In fact, looking
at Fig. 3 one
can expect the $\omega_{+}$ mode to have the worse convergence of all, as it
really happens in our calculation.
This agrees with TDLDA results \cite{13} that show how
the peaks in the excitation spectra are noticeably wider for the $\omega_{+}$
mode than for the $\omega_{-}$ mode.

The dependence of the dispersion relation on the width $d$ of the slab can be
seen in Fig. 4 (a)-4 (b). The $\omega_{-}$ and the $\omega_{+}$
modes tend to degenerate when $k>>d^{-1}$ and the same is true for the even and
the odd multipole modes. This is due to the fact that the symmetry of
the mode is not a real difference when the density fluctuation of
two modes is concentrated on the two surfaces of the slab and these are
enough apart in comparison with the wavelength of the mode. Obviously,
for larger values of $d$ plasmons will tend to degenerate for smaller
values of $k$. The $\omega_{-}$ and the $\omega_{+}$ modes degenerate
at an energy close to $\omega_{p}/\sqrt{2}$. In the semi-infinite
medium limit their dispersion relations will join at the origin and
they will form the well known
surface plasmon. The even and the
odd multipole modes will also join in a unique multipole plasmon
mode \cite{15} for extremely large slabs. In Fig. 4 (b) can be seen that
the dispersion relation of
the odd plasmon mode is much more affected by the change of the
width of the slab than that of the even multipole mode, so the dispersion
relation of the surface multipole mode will be closer to that
of this last mode.

In Fig. 5 (a) the dispersion relations of two slabs with a
different electronic density are plotted. The sodium slab has a greater
electronic density and the energy of its modes is higher than those
of the potassium slab. Surface modes degenerate for lower values
of $k$ in slabs of this last metal. The increase in the energy of
the modes in a high-density metal is due to both kinetic and coulomb
energies although in relative terms the coulomb term is more responsible
of this feature.
Moreover, the density also influences the diffuseness of
the electronic profile.
As discussed earlier, these two factors have an important effect
on the convergence of the expansion series. A plot of the plasmon
modes for an aluminium slab is displayed in Fig. 5 (b). Besides
the rise of the energies of the modes we can notice a greater gap between
the origin of the odd multipole mode and the first bulk resonance.
This is not due directly to the higher electronic density but to the
steeper slope of the electronic profile as will be seen soon.\\
In order to distinguish the effects due to the change of electronic
density and those caused by a different diffuseness of the profile we
have studied three systems where we have changed only the length of
the intermediate step. The results are represented in Fig. 6.
When the length of the step is reduced the origins of the $\omega_{+}$
and of the even multipole tend to rise, they are no longer at
$0.8\omega_{p}$, and a gap appears between them.
It is clear that the origin $0.8 \omega_p $ depends
strongly on the diffuseness of the surface and can
be used as a test on how good is an electronic profile.
The odd multipole and the first bulk resonance also have higher energies
for $k=0$ and the gap between their origins also increases. The slopes
of the dispersion relations of the even and the odd multipole modes suffer
a significant change and are closer than before. Finally, for a system
with no diffuseness, the origin of the $\omega_{+}$ mode has risen up to
$\omega_{p}$ and the two dispersion relations of the multipole modes
are in the same place as the first and second bulk resonances in the
previous systems. The result of the reduction of the
diffuseness is, then, the gradual conversion of the even and the odd
multipole modes into the first two bulk resonances.

\section{Summary and conclusions}

We have succeeded in developing a semi-classical calculation that
includes as solutions
the even and the odd multipole 
modes, that are not usually considered in
related calculations. In the semi-infinite medium they are brought
together into the by now well studied multipole plasmon mode.
These multipole modes modify the classical origin of the
$\omega_{+}$ mode and are closely 
dependent on the diffuseness of the
electronic profile. Their origin is shifted upwards when the
diffuseness is reduced and they disappear when it is set to zero.

The $\omega_{-}$ plasmon is far more observable, for intermediate
values of $k$, than the $\omega{+}$ plasmon. This can be stated
either examining the convergence of the results on the number of
terms of the expansion series or taking into account the relative
contribution of the coulomb and the kinetic energies. The bulk
resonances closer to $\omega_{p}$ are also realistically described.
The Thomas-Fermi density functional gives a too flat slope of the
kinetic energy. The exchange and correlation contribution lowers
the energy for large values of $k$.

Local RPA is a useful and economic method for finding the
excitation operator $Q$ whenever well-defined collective
excitations are expected. The size of the system determines, among other
factors, the number of terms needed in the expansion series.
If a strong damping acts on the plasmon modes due to high
diffuseness of the electronic profile, low electronic density or
whatever other cause, the convergence can be difficult to reach or even
non-existent at all.

\vspace{5mm}

\noindent{\large \bf Acknowledgments}

We thank N. Barberan and D. Espriu for a critical reading
of the manuscript. S.L. and A.T. acknowledge a grant from
the {\it Generalitat de Catalunya}. This work has been partially supported
by DGICYT project PB93-0035 and
CICYT grant AEN93-0695 and CIRIT contract GRQ93-1047.

\newpage

\appendix

\section{Appendix}

In this appendix we will present the formulae 
which were used in the solution of 
equation \ref{eigen}.

The first object to compute, that is $\cal B$, is
a straightforward calculation;
\begin{equation}
{\cal B}_{(k,p),(k',p')}=A(k^2+pp')\int{dz \exp{\left(-i(p-p')z\right)}
n(z)}
\end{equation}

To compute $\cal K_{\alpha \beta}$ we start by considering;
\begin{eqnarray}
{\cal M_{\alpha \beta}} &\equiv & {\cal M_{\alpha \beta}}^k+
{\cal M_{\alpha \beta}}^{ex}+{\cal M_{\alpha \beta}}^{c}
\nonumber
\\
&=&\langle \Phi_0 |\exp{\eta \left( \left[Q_{\alpha},H\right] \right)
}
\exp{\left(
\eta'\left[Q_{\beta},H\right]
\right) }\left( H_k+H_{ex}+H_c\right)\times \nonumber
\\&&
\exp{\left( -\eta'\left[Q_{\beta},H\right] \right) } \exp{\left( -\eta
\left[Q_{\alpha},H\right] \right) }|\Phi_0 \rangle
\nonumber
\\
&\equiv &\langle \eta \eta'|H_k+H_{ex}+H_c|\eta \eta'\rangle
\end{eqnarray}
The previous object is useful as it is connected to the one we are
interested in through formulae;

\begin{equation}
{\cal K}_{\alpha \beta}= \left( \frac{\partial ^2}{\partial \eta
\partial \eta' } {\cal M}_{\alpha \beta} \right)_{\eta= 0,\eta'= 0}
\end{equation}

At this point we have reduced the problem of calculating the expressions
in formula \ref{calK} to the evaluation of expressions such as
$\langle \eta \eta'|F(n)|\eta \eta'\rangle$
which are still quite intractable. Some approximations are
needed to go any further. We will approximate terms like this one as
\begin{equation}
\langle \eta \eta'|F(n)|\eta \eta'\rangle=
F(\langle \eta \eta'|n|\eta \eta'\rangle)+...
\end{equation}
and keep just the first term.

The rest of the computation 
is straightforward but lengthy. We arrive at
\begin{eqnarray}
{\cal K}_{(k,p),(k,p')}^{c}&=&
4 \pi A p p'\int{dz n(z)^2 \exp(-i(p-p')z)}
\nonumber
\\
&&+2\pi A \int dz dz' n(z) n(z') \exp(-i(p-p')z)\exp(-|z-z'| k)\times
\nonumber
\\
&&\left( k^3-pp'k+i(p+p')k^2sgn(z-z') \right)
\nonumber
\\
&&-4\pi A p p'\int{dz \left(n(z)-n_j(z') \right)n(z)\exp(-i(p-p')z)}
\nonumber
\\
&&-2i\pi A \left(pp'+k^2 \right)p' \int dzdz' \left( n(z)-n_j(z')
\right) n(z) \exp(-iz(p-p')z) sgn(z-z')
\nonumber
\\
&&
\end{eqnarray}
The exchange term is easily computed with the result;
\begin{eqnarray}
{\cal K}_{(k,p),(k,p')}^{ex}&=&
A(k^2+p^2)(k^2+{p'}^2) \int dz n(z)^2
\frac{\partial^2\epsilon_x(n)}{\partial n(z)^2}
\nonumber
\\
&&-(k^2+{p'}^2)p(p-p') \int dz \exp(-i(p-p')z) \left( n(z)
\frac{\partial \epsilon_x(n)}{\partial n(z)}-\epsilon_x(n) \right)
\nonumber
\\
&&
\end{eqnarray}
The kinetic part does not show up any further complication as compared
to the exchange term since where one reads $\epsilon_{ex}$ one
substitutes $\tau(n)$ and the rest is exactly the same.

\vfill
\eject

\newpage

{\large \bf Figure captions}

\vspace{0.5cm}

{\bf Figure 1 (a)}: Dispersion relation of the plasmon modes below the plasma
frequency in a sodium slab ($d=25 a_{0}$ where $a_{0}$ is the Bohr
radius).
From bottom to top can be seen the $\omega_{-}$,
$\omega_{+}$,even multipole and odd multipole modes. Thin lines
represent, from bottom to top, $\omega_{p}/\sqrt{2}$, $0.8\omega_{p}$
and $\omega_{p}$.

{\bf Figure 1 (b)}: Dispersion relation of the four lowest bulk plasmon
resonances in
the same case as Fig. 1 (a). The thin line represents the plasma
frequency.

{\bf Figure 2 (a)}: Sudden approximation to the density fluctuation calculated
as explained in the text for a sodium slab ($d=40 a_{0}$). The dashed
line represents the $\omega_{-}$ mode while the solid line is the
$\omega_{+}$ mode.

{\bf Figure 2 (b)}: The dashed line represents the density fluctuation of the
even multipole mode in the same case as Fig. 2 (a). The solid
line is the odd multipole mode.

{\bf Figure 2 (c)}: Density fluctuation of the two lowest bulk plasmon
resonances for
the same system as Fig. 2 (a). The dashed line represents the lowest
resonance and the solid line represents the next resonance.

{\bf Figure 2 (d)}: The dashed line is the third bulk resonance with lowest
energy and the solid line is the fourth. Calculation for the same
system as previous cases.

{\bf Figure 3}: The energy contributions of the same mode are plotted in
the same type of line. The upper one is the coulomb contribution and
the lower one is the kinetic contribution. The solid line
corresponds to the $\omega_{-}$ mode, the long-dashed to the
$\omega{+}$ mode, the short-dashed to the even multipole mode,
the dotted line to the odd multipole mode and the dot-dashed line
corresponds to the lowest bulk resonance. The calculations are
made for the same case as Fig. 1 (a).

{\bf Figure 4 (a)}: Dispersion relation of the $\omega_{-}$ and the $\omega_{+}$
modes for a sodium slabs of different widths. The solid line represents
a width $18 a_{0}$, the long-dashed line stands for $25 a_{0}$ and
the short-dashed line is the result for $40 a_{0}$.

{\bf Figure 4 (b)}: Same as Fig. 4 (a) but with the even and the odd
multipole plasmon modes.

{\bf Figure 5 (a)}: Dispersion relation of the plasmon
modes of two slabs of different metals and the
same width ($d=25 a_{0}$). The solid line
corresponds to a potassium slab. The dashed
line is used for the sodium slab.

{\bf Figure 5 (b)}: Dispersion relation of the plasmon
modes of an aluminium slab. The width is $25 a_{0}$.

{\bf Figure 6}: Dispersion relation of the
four modes with lowest energy. The solid line represents
the modes with an intermediate step set to $\delta=2.11 a_{0}$.
The long-dashed lines are calculated with $\delta=1 a_{0}$ while
while short-dashed lines stand for a system with $\delta=0$.
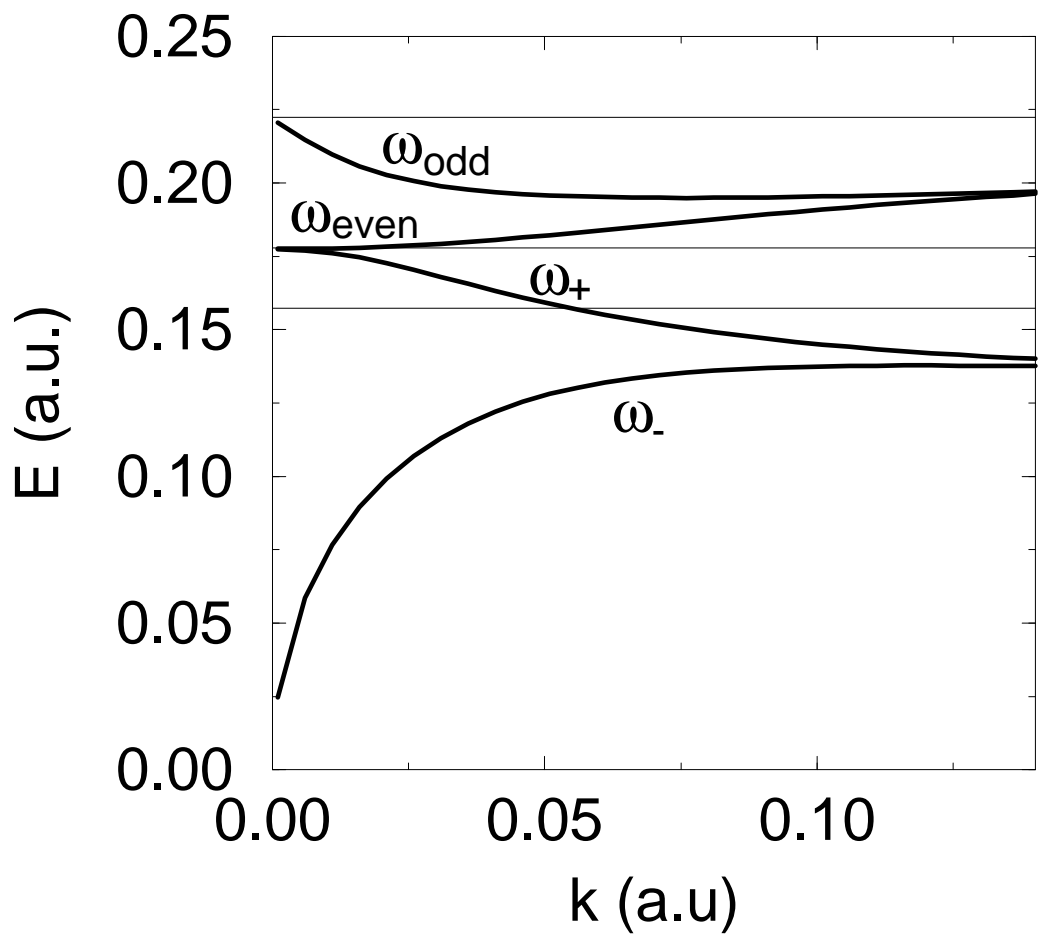
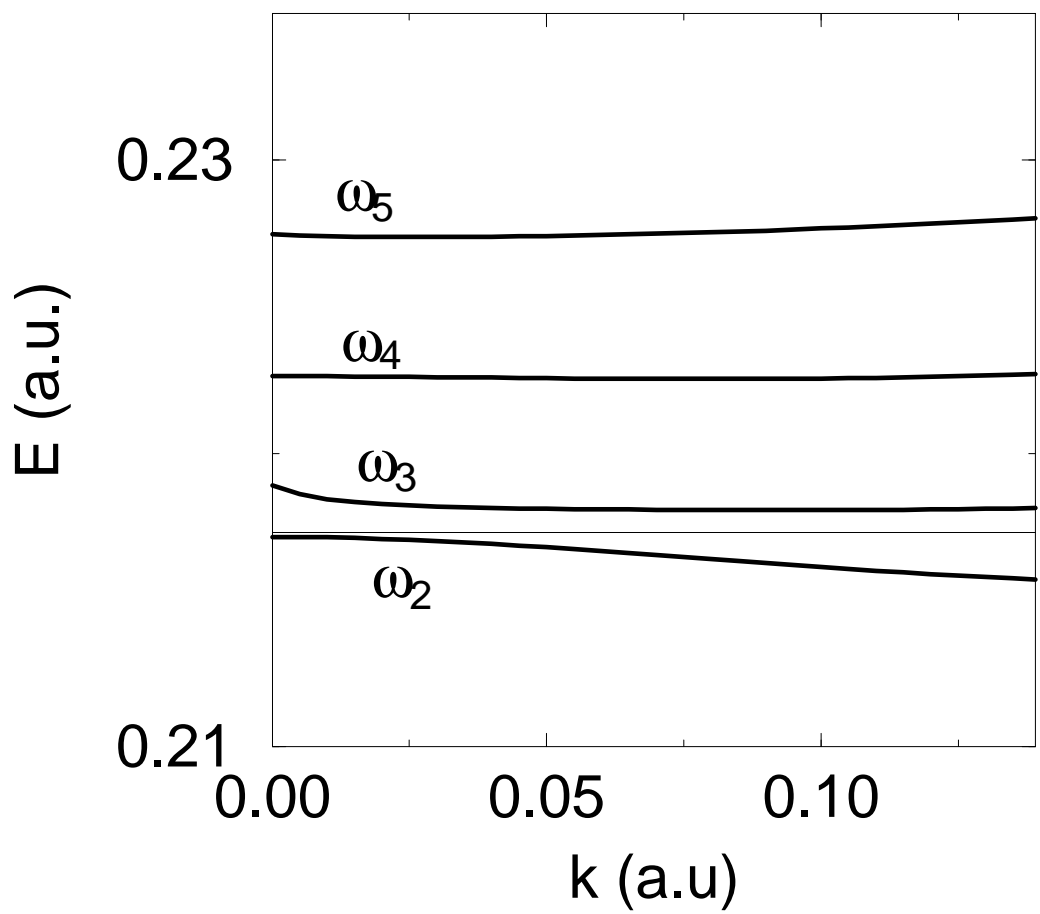
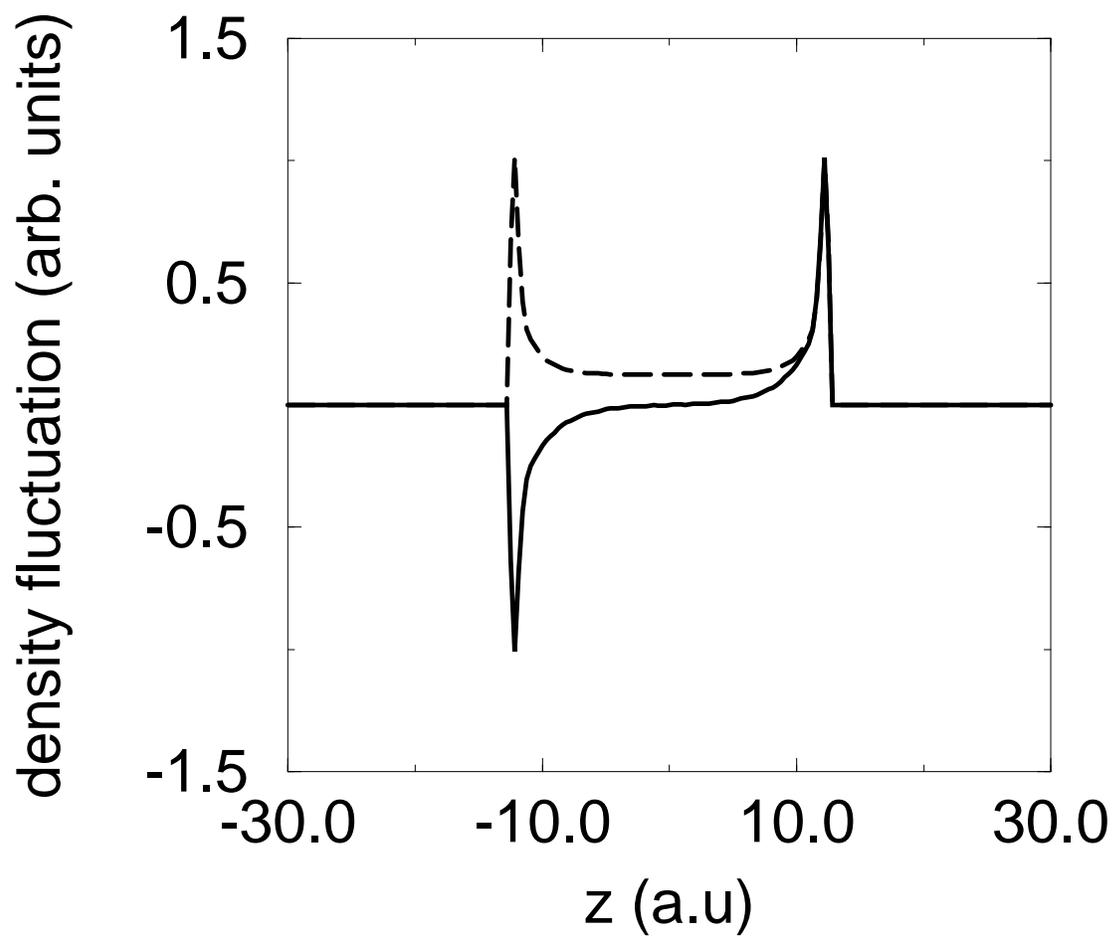
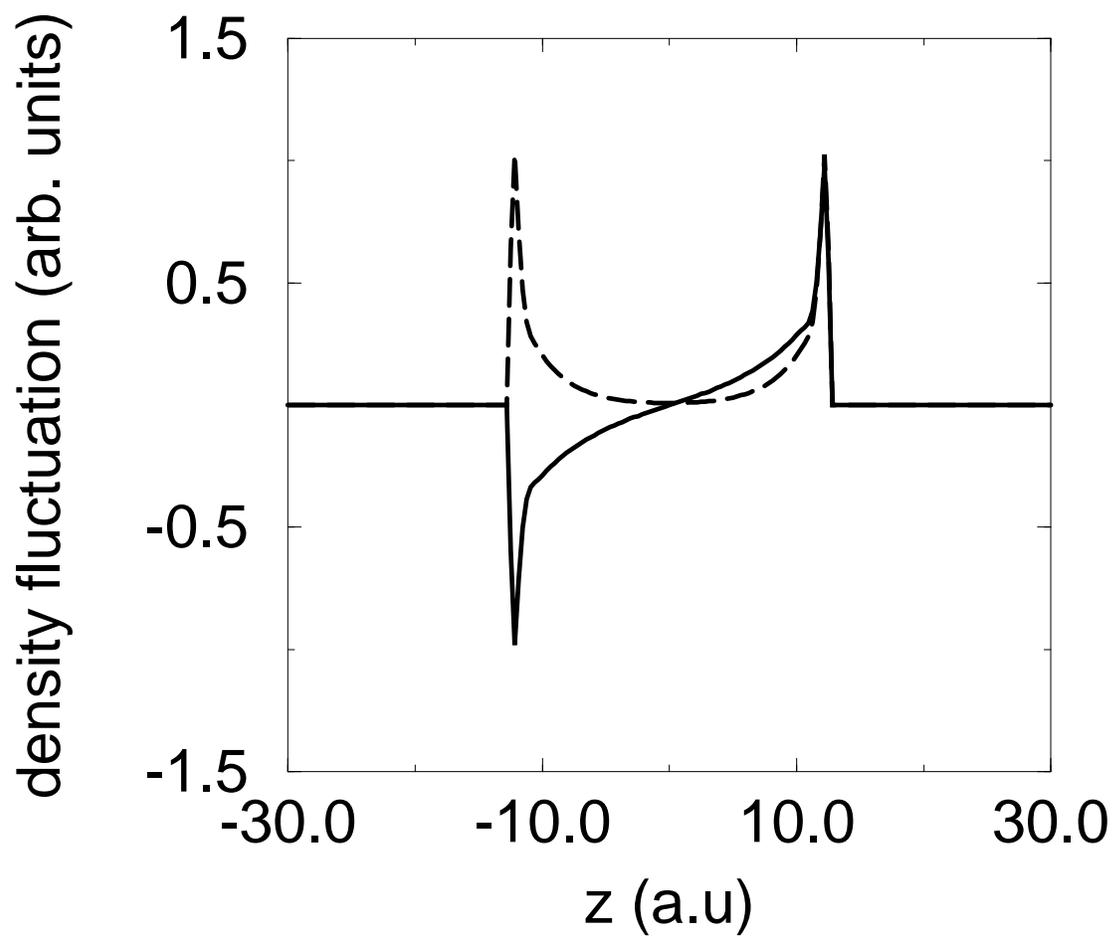
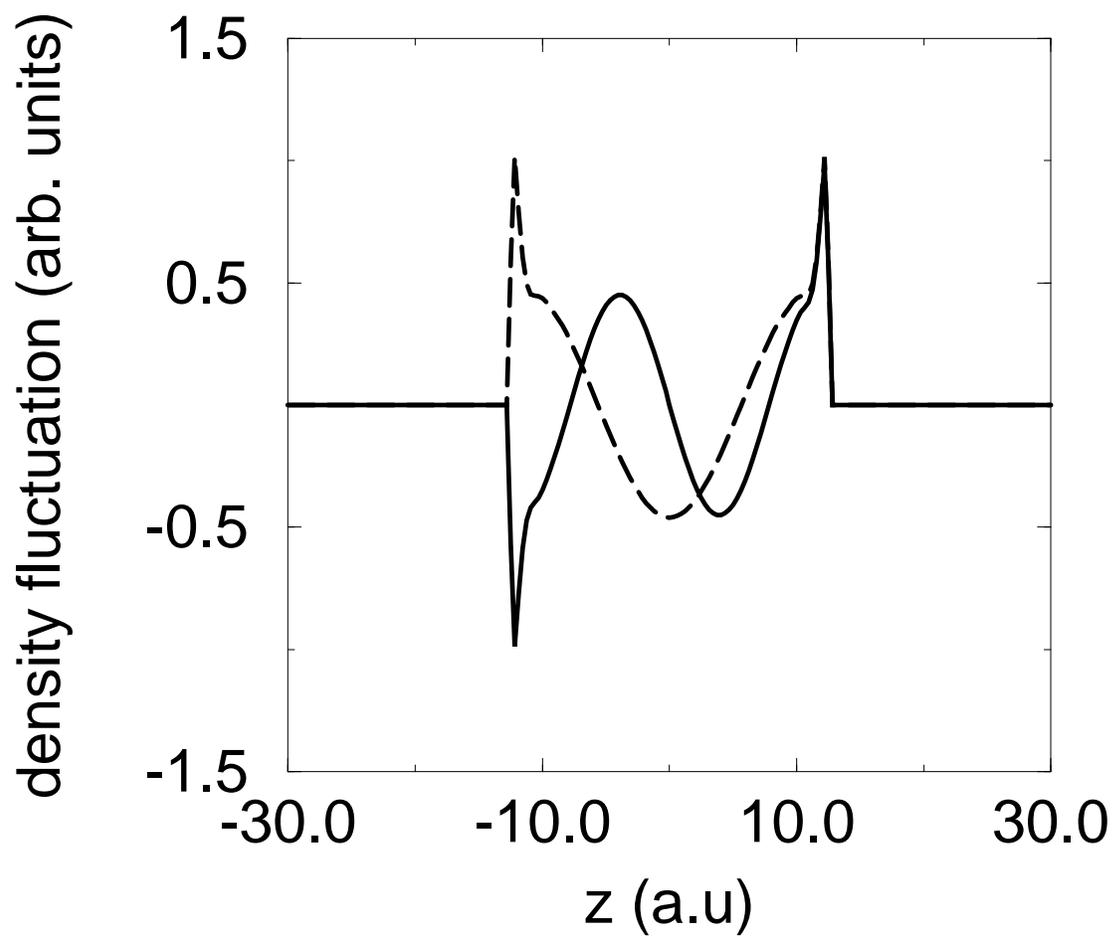
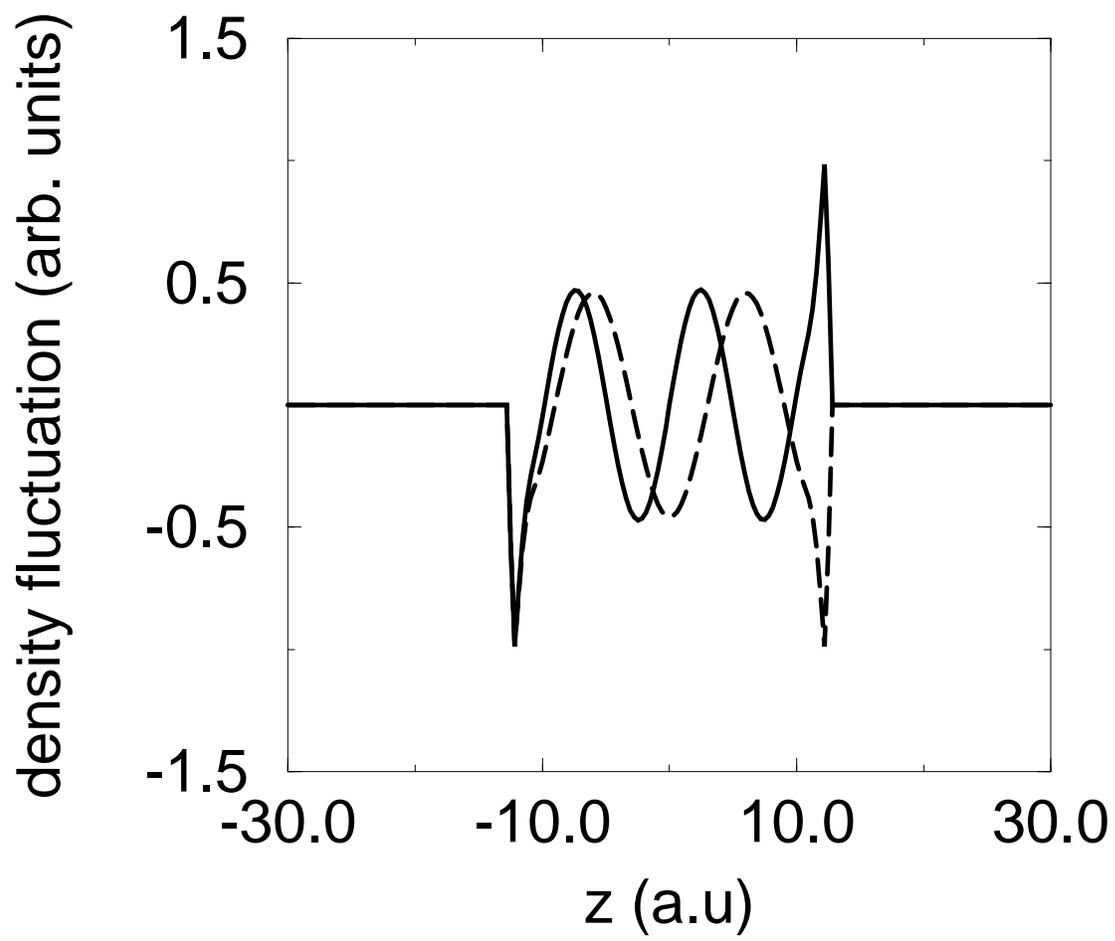
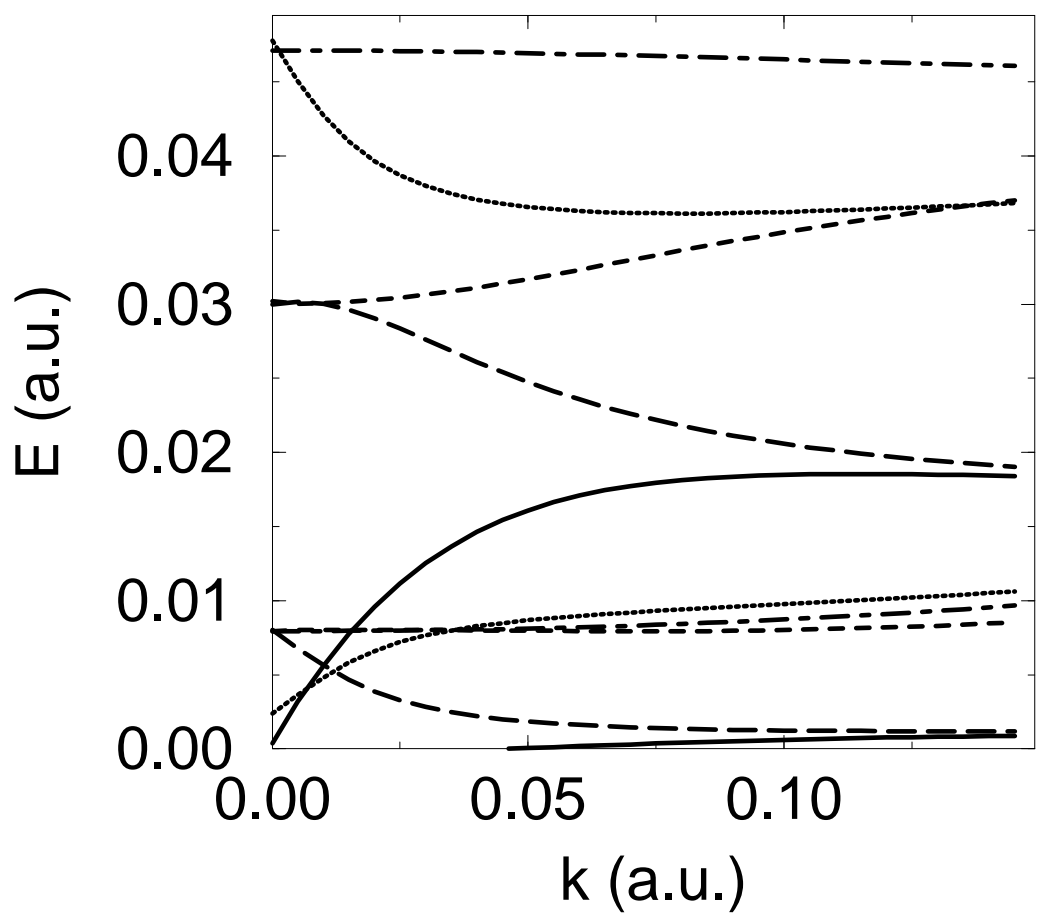
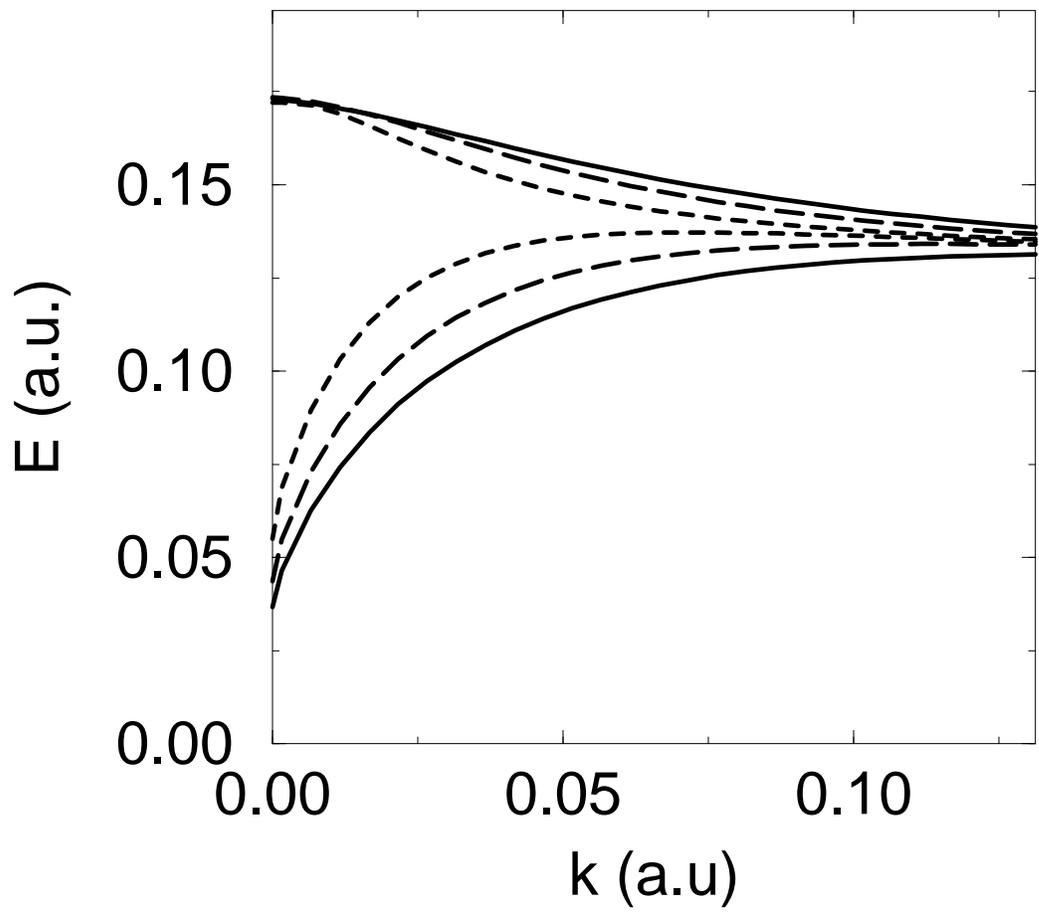
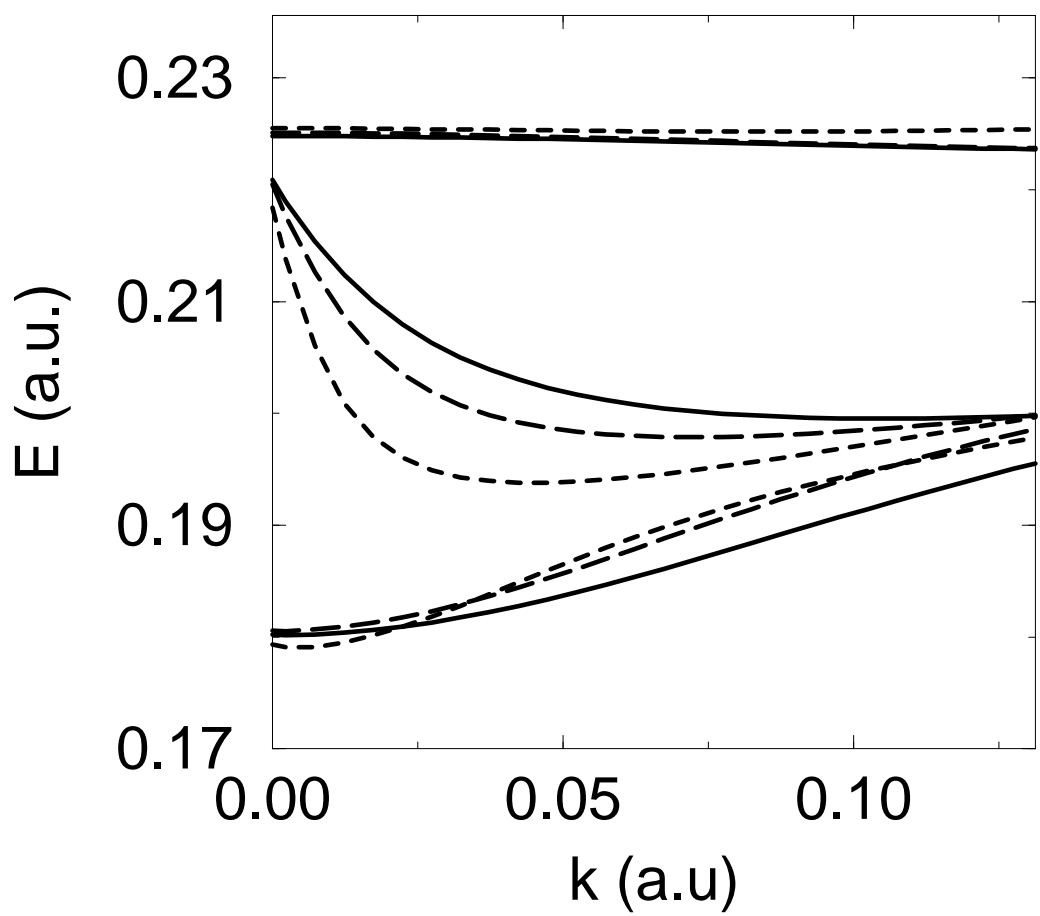
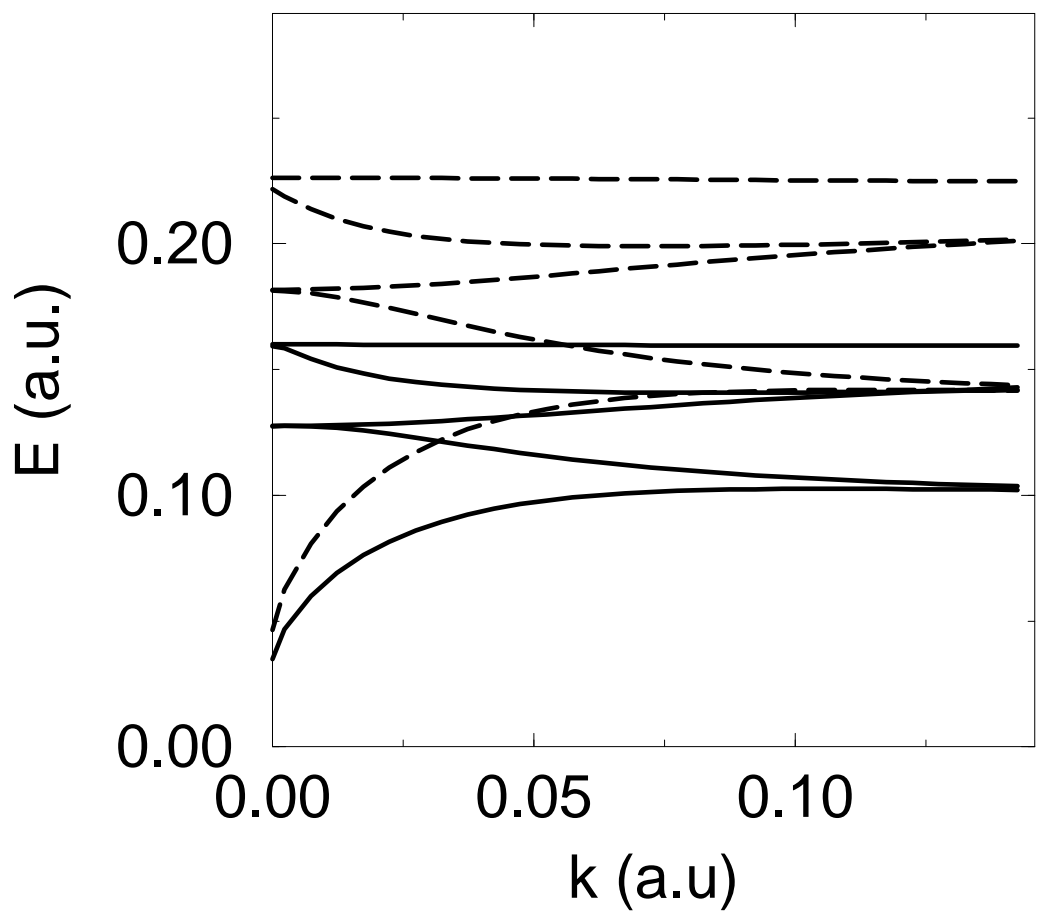
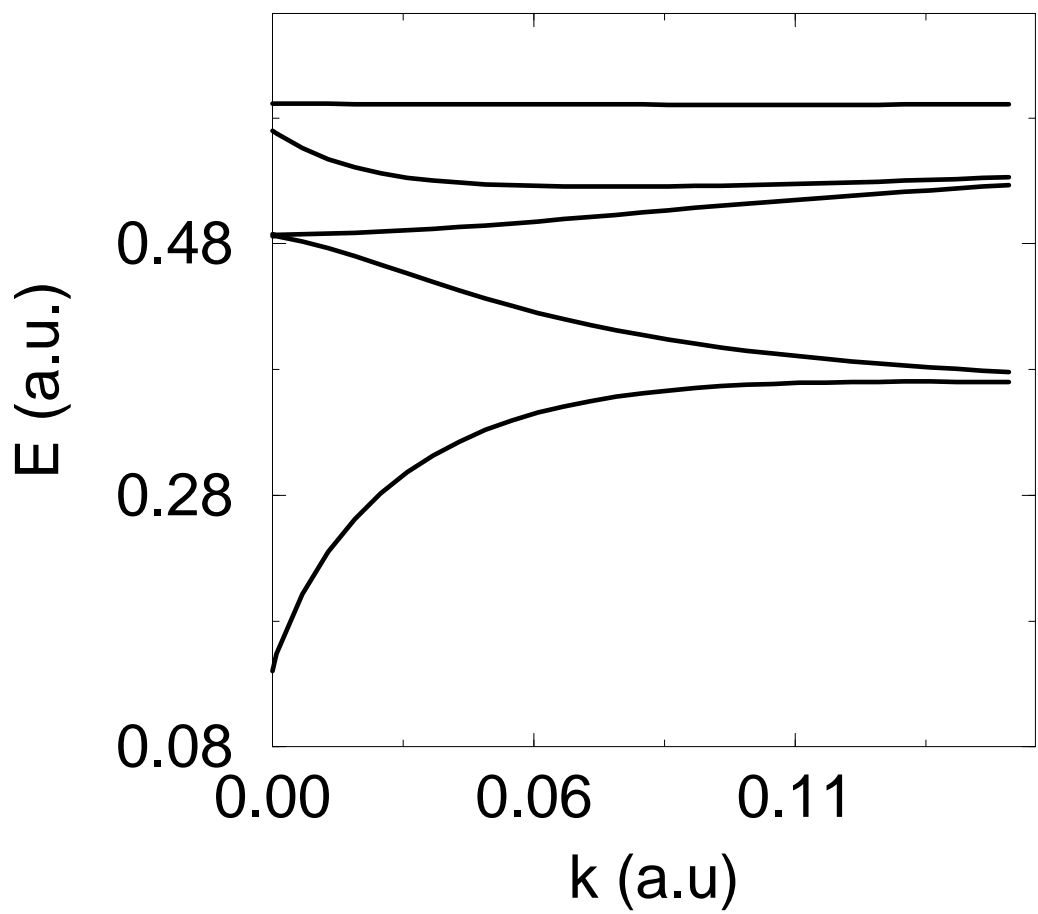
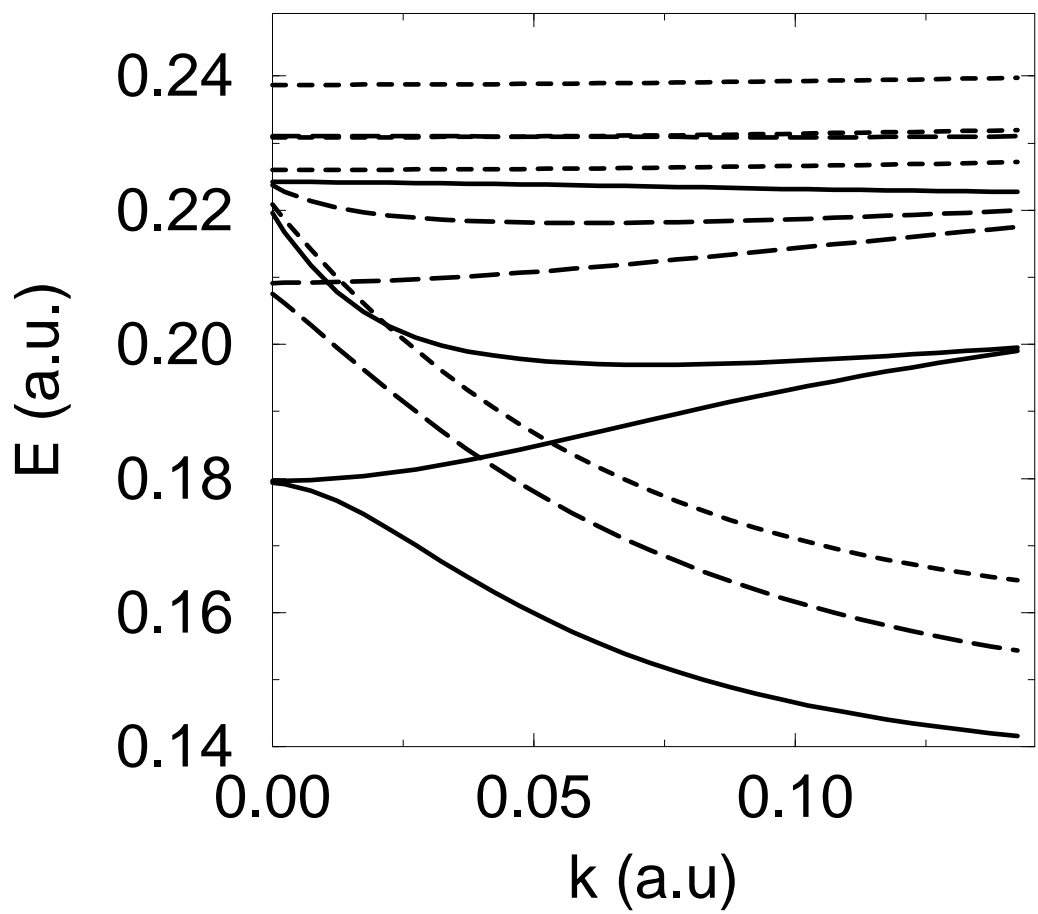
\end{document}